# The Maxwell Demon and Market Efficiency

Roger D. Jones, Sven G. Redsun, Roger E. Frye, & Kelly D. Myers

*CommodiCast, Inc. & Complexica, Inc., 125 Lincoln Avenue, Suite 400, Santa Fe, NM 87501 USA*

**This paper addresses two seemingly unrelated problems, (a) What is the entropy and energy accounting in the Maxwell Demon problem? and (b) How can the efficiency of markets be measured? Here we show, in a simple model for the Maxwell Demon, the entropy of the universe increases by an amount $\eta \approx 0.8399955201358$ bits/particle in going from a random state to an ordered state and by an amount $\eta^* \approx 2.373138220832$ bits/particle in going from one sorted state to another sorted state. We calculate the efficiency of an engine driven by the Maxwell sorting process. The efficiency depends only on the temperatures of the particles and of the computer the Demon uses to sort the particles. We also show the approach is general and create a simple model of a stock market in which the Limit Trader plays the role of the Maxwell Demon. We use this model to define and measure market efficiency.**

Maxwell[1][2][3] created his Demon in 1867 to help clarify the issues associated with the Second Law of Thermodynamics. In particular, Maxwell wished to address the question of the role of intelligence in the flow of entropy. The Demon was an intelligent microscopic creature that sat at a trapdoor separating a box into two sides. Particles inhabited both sides of the box. The Demon observed the particles and allowed fast particles to enter into one side of the box and slow particles to enter into the other side of the box. The entropy of the particles was thus decreased and a temperature gradient, capable of producing useful work, was created. The intelligent Demon seemed to violate the Second Law. Either the Second Law had to be abandoned or the entropy of the Demon had to increase to compensate for the decrease in entropy of the particles. Szilard[4] was the first to note that there must be entropy associated with information. Shannon[5] developed an entropy measure for information that mirrored the logarithmic entropy measure of Boltzmann in physics. Brillouin[6] unsuccessfully tried to destroy the Demon by concentrating on the entropy of the measurement process.

The problem of market efficiency is more modern. The Efficient Market Hypothesis[7] (EMH) states that, at any given time, prices reflect all available information available about a particular stock and/or market. According to EMH, no trader has an advantage in predicting the return of a stock since every trader has access to all information. Prices are commonly thought to follow a "random walk[8]." The EMH comes in three flavours:

**Strong Efficiency** – All information in the market, public or private, is accounted for in the stock price. Even insider knowledge cannot provide price predictability.

**Semi-Strong Efficiency** – Public information is accounted for in the stock price, but not insider information. Neither analysis of the company fundamentals nor technical analysis of the history of the price/volumes can be used to predict prices.

**Weak Efficiency** – All information in historical stock prices is reflected in today's price. Technical analysis of the history cannot be used to predict future prices.

There have been extensive measurements of market efficiencies[9][10] with the typical conclusion that markets are efficient and prices are unpredictable, at least in the weak sense. Of course, markets cannot be efficient on all timescales. There is some small finite time in which information propagates through the market. The approach to efficiency on these small timescales has also been measured.[11]

We take a much different approach to understanding the information content of a market. Rather than price, we focus on the number of shares traded or offered for trade. This information is not independent of price information. Price and trade volume are related through liquidity or price impact.[12] The number of shares traded, however, is more directly related to supply/demand and also to the probability measures of information theory. In particular, we use the theory developed in the solution of the Maxwell Demon Problem to develop a theory of an ideal stock market. When the behaviour of stocks in actual markets is compared with the behaviour of the idealized market we are led to the conclusion that the market does not appear to be a *Poisson* process throughout the day. In fact, for part of the trading day, the market appears to be responsive to the supply/demand for market orders.

In the Maxwell Demon Problem we have an intelligent gnome that orders a box of particles so that hot particles are on one side of the box and cold particles are on the other. In a stock market we have two types of traders, Market Traders and Limit Traders. Market Traders place orders and take whatever price the market provides. In a market composed solely of Market Traders chaos would reign because no mechanism would exist for determining a price. Limit Traders are traders that place orders to be filled at a given price. The Limit Traders therefore establish a price in a market. They play the role that Maxwell's Demon plays in ordering a box of particles. Limit Traders order the market and give meaning to concept of price. The arrival of market orders on the trading floor is analogous to the arrival of a particle at a trapdoor between the two sides of Maxwell's box. The decision by the Demon to open the trapdoor for the particle or close it is analogous to the decision by the Limit Trader to go long or go short in anticipation of the market order. Although the two problems do not map exactly onto each other, the same excess entropy constants emerge in each problem.

Our goal in addressing the Maxwell Demon question is to design a theoretical engine that generates useful work from the intelligent sorting of particles and to use that engine to account for energy and entropy flows, including flows into and out of any computer the Demon uses to help sort the particles. In order for the Second Law to hold in our idealized world, the engine must have efficiency less than one and the total entropy of the universe must increase as the engine runs. Fortunately, we find this to be the case. The efficiency of the engine is determined by the temperature of the particles and the temperature of the Demon's computer.

Our goal in the market problem is to develop a measure of the efficiency of information use in a market. We would then like to use this measure to calculate the efficiencies of trading of any particular stock. From a sociological point of view this provides a window into market behaviour. From a business point of view, this measure can be used to inform trading decisions.

**The Maxwell Demon**

Our version of the Maxwell Demon problem is illustrated in Figure 1. The Maxwell Demon observes the particles and writes the relevant information on the next particle to approach the barrier to an Input Tape. The Computer writes the appropriate control signals to an Output Tape, or Controller. The Controller then sends the appropriate signals to the barrier to open or close to allow the particle to pass or not to pass. The Computer and its Input/Output are immersed in a heat bath. We neglect any work or entropy generation associated with opening and closing the barrier, with measuring the state of the particles, and with any computation other than writing and erasing tapes. We do, however, take into account the work and entropy generation associated with writing and erasing tapes and entropy/work associated with the particles.

Zurek pointed out[13] that the total entropy, which he calls *physical entropy,* must include both the usual *statistical entropy* of the system as well as the entropy of the computer (including input and output tapes), which he calls *algorithmic information*. The *algorithmic information* is the most concise message which describes the system with the requisite accuracy.

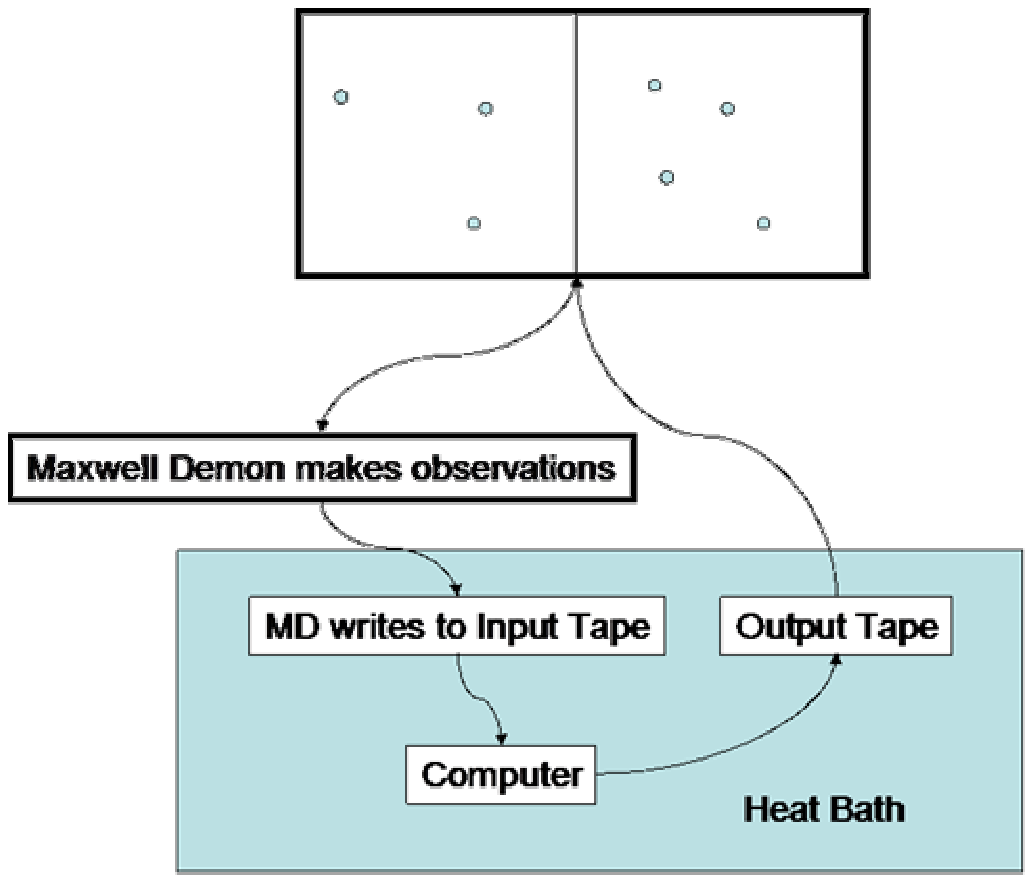

Figure 1: The Maxwell Demon observes the particles and writes relevant information to an Input Tape that is immersed in a heat bath. The Input Tape is processed by the Computer also immersed in the heat bath. The Computer writes to an Output Tape, or Controller, that sends control signals to a barrier that allows particles to pass or not.

In the case in which particles start out randomly distributed throughout the box (unsorted), the time dependence of the fraction of $N$ particles of type $A$ or $B$ being on the right ($R$) or left ($L$) side of the box is

$$p_t(A,L) = p_t(B,R) = \frac{1}{2} - p_t(A,R) = \frac{1}{2} - p_t(B,L) = \frac{1}{2}(1 - \Pi_t) \text{ where } \Pi_t \equiv \frac{1}{2}\left(1 - \frac{1}{N}\right)^t.$$

These fractions are the probabilities that a particle of a given type on a given side of the box approaches the barrier at time $t$. The entropy of the Controller is then calculated to be $\Delta S_{\text{controller}} = N + \eta N$ bits where $\eta = \sum_{t=0}^{\infty}\left\{-\frac{1}{N}\Pi_t \lg[\Pi_t] - \frac{1}{N}(1 - \Pi_t)\lg[1 - \Pi_t]\right\} - 1 = \frac{\pi^2}{12\ln(2)} - \frac{\ln(2)}{2}$ bits.

In the case in which the particles initially are sorted, but on the wrong side of the box (anti-sorted), the probabilities are

$$p_t(A,L) = p_t(B,R) = \frac{1}{2} - p_t(A,R) = \frac{1}{2} - p_t(B,L) = \left(\frac{1}{2} - \Pi_t\right),$$ and the entropy of the

Controller is $\Delta S_{\text{controller}}^* = \eta^* N$ where

$$\eta^* = \sum_{t=0}^{\infty}\left\{-\frac{1}{N}2\Pi_t \lg[2\Pi_t] - \frac{1}{N}(1 - 2\Pi_t)\lg[1 - 2\Pi_t]\right\} = \frac{\pi^2}{6\ln(2)} = 2\eta + \ln(2) \text{ bits. This}$$

constant also appears in one dimensional chaotic time series[14] as the Liapunov exponent for the Gauss map. Here, the quantities $\delta S \equiv \eta N$ and $\delta S^* = \eta^* N$ are known as *excess entropies*. These entropies represent the irreversible entropy increase of the universe due to the sorting process.

For each of these initial conditions the Demon collects an extra bit of information per particle approach that is not used in the control of the barrier. We call this unused information *maintenance information*.

Up to this point the results are based on information theoretic arguments, independent of physics. Energy accounting allows us to tie the problem to physics.

Once the particles are separated, the system of particles is available to do useful work. Therefore, we can create an engine that is driven by the "Intellectual Property" of the Maxwell Demon (Figure 2). The engine is coupled to a computer that has as its basic storage element a cylinder immersed in a heat bath of temperature $T_S$ and composed of a single particle and two pistons. The computational element is illustrated in Figure 3. The Demon writes the information he collects to an Input Tape composed of computation units like that illustrated in Figure 3. The Controller is also composed of the same computation units.

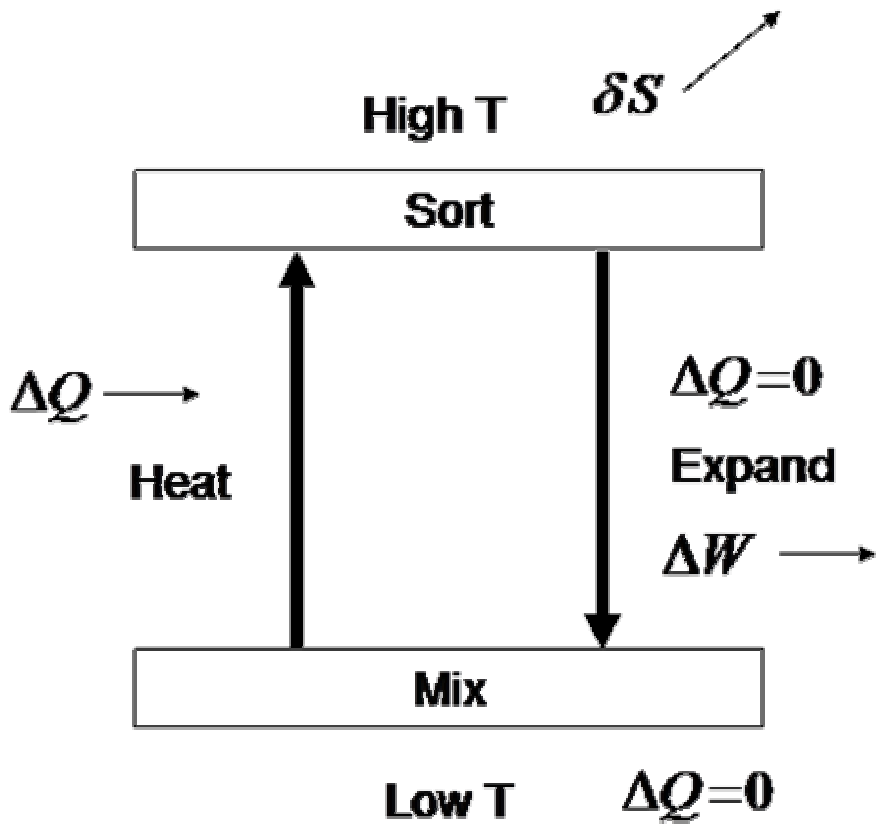

Figure 2: The Bistro Drive. The engine experiences four phases, a sorting phase, a work phase, a mixing phase, and a heating phase. The sorting phase separates the hot particles from the cold particles and generates heat $\Delta Q_{\text{sort}} = N k_B T_S \ln(2) \eta = k_B T \ln(2) \delta S$ where $k_B$ is Boltzmann's Constant. In the work phase, the piston expands adiabatically against the cold gas until pressure

equilibrium is attained. There is no heat transfer in this phase, although the gas cools. The piston is removed in the mixing phase allowing the particles to access the entire box. The gas temperature remains constant in this phase. Finally, the gas is reheated to its original temperature.

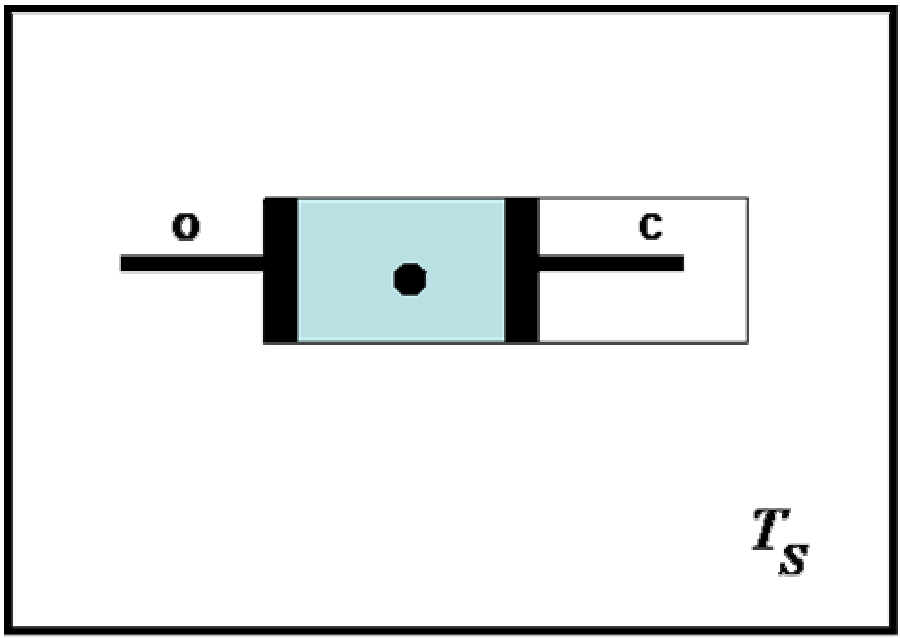

Figure 3: A Controller Memory Unit. This unit is signalling for the barrier to close. The c piston is in the middle of the cylinder. This occurs when the Input Tape is indicating that a slow particle is approaching from the right or a fast particle is approaching from the left. If the Input Tape is signalling for an open barrier then the o piston in the Controller Unit is in the middle of the container. A Unit that is erased has both pistons at the end of the container.

The efficiency of this engine for a monotonic ideal gas is then $\Upsilon = \dfrac{\dfrac{3}{2}T_p\chi}{\dfrac{3}{2}T_p\chi + T_S\ln(2)(1+\eta)}$, $\chi \approx 0.0777$ where $T_p$ is the temperature of the particles and $T_S$ is the temperature of the heat bath surrounding the computer. Like a Carnot Engine, the efficiency depends on two temperatures. The lower temperature, however, is that of the heat bath surrounding the computer.

**Market Efficiency**

We now apply the results of the Maxwell Demon problem to a simple artificial stock market. This will allow us to define an idealized market that can be compared with real markets. We consider a very simple model for the stock market that includes a supply and demand of stock through short and long market orders and a single Limit Trader who fills all market orders.

We imagine we have $N$ market orders, some of which are unfilled long orders, some of which are unfilled short orders, and some of which are empty orders (a time period in which there is no market order placed). As time progresses all unfilled orders become filled orders. Unfilled orders become filled by being paired with limit orders. We imagine a Limit Trader who is cognizant of the type of the next market order and who has placed a limit order in anticipation of the market order. The Limit Trader is a fully prescient inside trader. This idealized Limit Trader plays the role of the Maxwell Demon for the System. This trader drives the market to order in the same way that the Maxwell Demon drives the box of unsorted particles to order. The Limit Trader anticipates the market orders and places an ask (a), a bid (b), or no order (0) in anticipation of short (S), long (L), or already filled (F) orders.

In the stock market model there are three states for the entities (S,L,F) rather than four, (AR, AL, BR, BL), in the Maxwell Demon model. There are three control actions (a,b,0) rather than two, closed or open barrier (c,o). Finally, the Limit Trader does not seem to collect Maintenance Information, while the Maxwell Demon does. The *excess entropy,* however, in the stock market model is exactly the same as in the Maxwell Demon problem. In particular, the *excess entropy* in going from a state in which all market orders are unfilled to one in which they are filled is $\delta S^* = \eta^* N$ , which is exactly the same *excess entropy* we calculated in going from a sorted state to a resorted state in the Maxwell Demon problem.

We define the *information efficiency* as $\hat{E}_t^* = 2^{\frac{\hat{S}_t - \hat{S}_f}{N}}$ where $\hat{S}_t$ and $\hat{S}_f$ are the measured entropy at time $t$ and at the end of the day, respectively. We measured the efficiency as a function of time during the trading day for two high volume stocks, Microsoft (MSFT) and Intel (INTC); two medium volume stocks, Network Appliance (NTAP) and PeopleSoft (PSFT); and two low volume stocks, C. H. Robinson Worldwide (CHRW) and Henry Schein, Inc. (HSIC) over the period August 1, 2003 to August 12, 2003. All six stocks are traded on the NASDAQ Stock Exchange. In that period there were approximately 1 million trades for a total of approximately 1.5 billion shares traded in the six stocks. Microsoft and Intel dominated the trade volume. Data were downloaded daily from S&P Comstock.

The fraction of orders filled, or *fill efficiency,* for Microsoft is displayed in Figure 4. Note that two regimes are apparent, a late-time regime in which the efficiency scales linearly with time consistent with a *Poisson* Process, and an early-time regime in which the efficiency is better described by a Supply/Demand Process. We see similar behaviour for all the stocks studied. The *information efficiency* is displayed in Figure 5.

While there is some variation in the efficiency during the trading day, the entropy for each end of day is within a few percent of $\left(1+\eta^{*}\right)N$ as prescribed by the theory.

Microsoft, for instance, has a mean entropy of 3.43 and a standard deviation of 0.15 as compared with the theoretical prediction of $1+\eta^{*} \approx 3.37$.

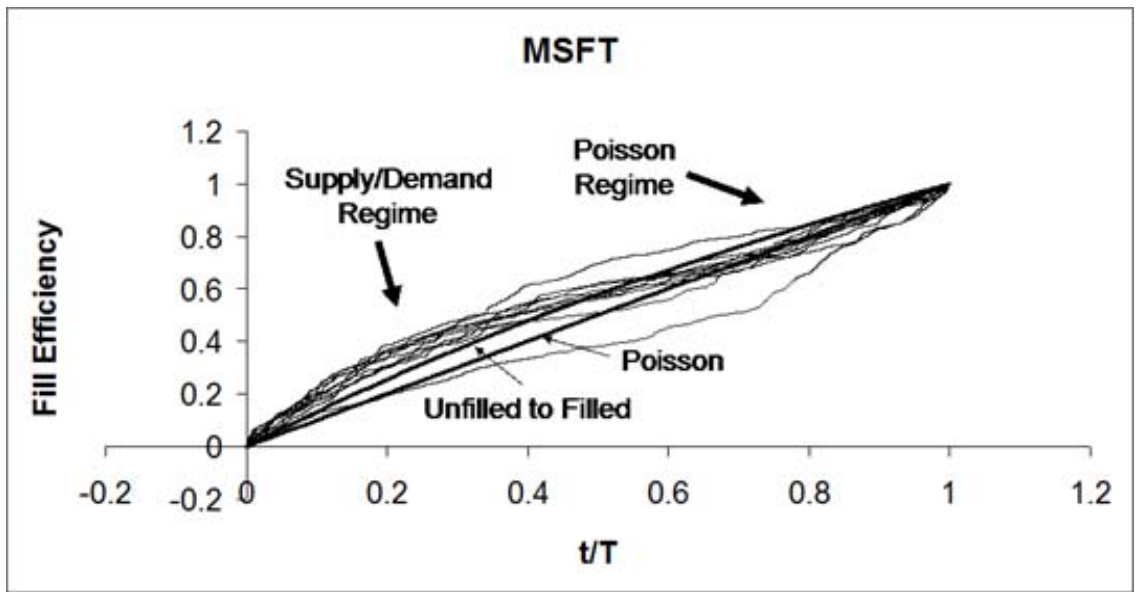

Figure 4: The Fill Efficiency vs. Time. Here, $T$ is the time from beginning to end of the trading day. Two theoretical curves are displayed, a Poisson curve that is proportional to time and a Supply/Demand curve that accounts for depletion of orders. We see two regimes. At late times the Fill Efficiency scales linearly with time as one would expect from a Poisson Process. At early times, the Fill Efficiency is better described by the Supply/Demand curve calculated from the market entropy.

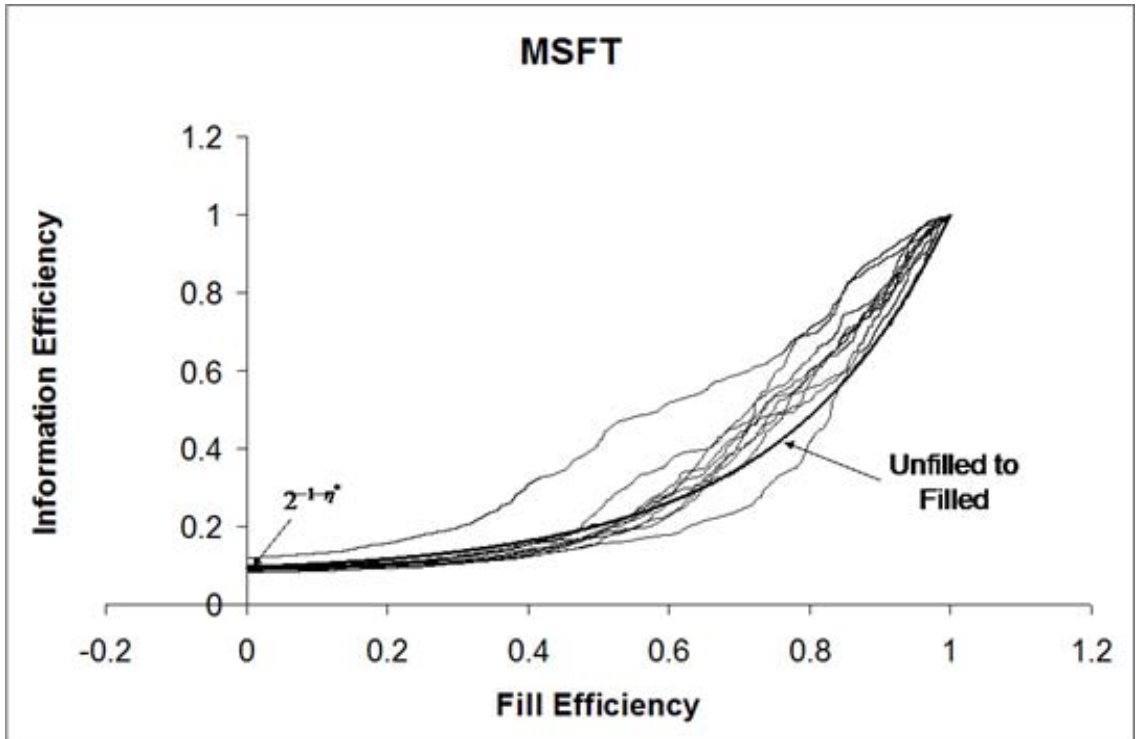

Figure 5: Information Efficiency measurements for Microsoft, a high volume stock. The theoretical forecast is displayed as a heavy line. The measured information efficiencies at zero fill efficiency are within a few percent of the theoretical predictions.

The *information efficiency* for a medium size stock, PeopleSoft, is displayed in Figure 6. The variation in entropy is the highest of the stocks we studied. This variation is due to several larger than average trades. Typically, larger than average trades drive the entropy to larger efficiency values than predicted by the theory.

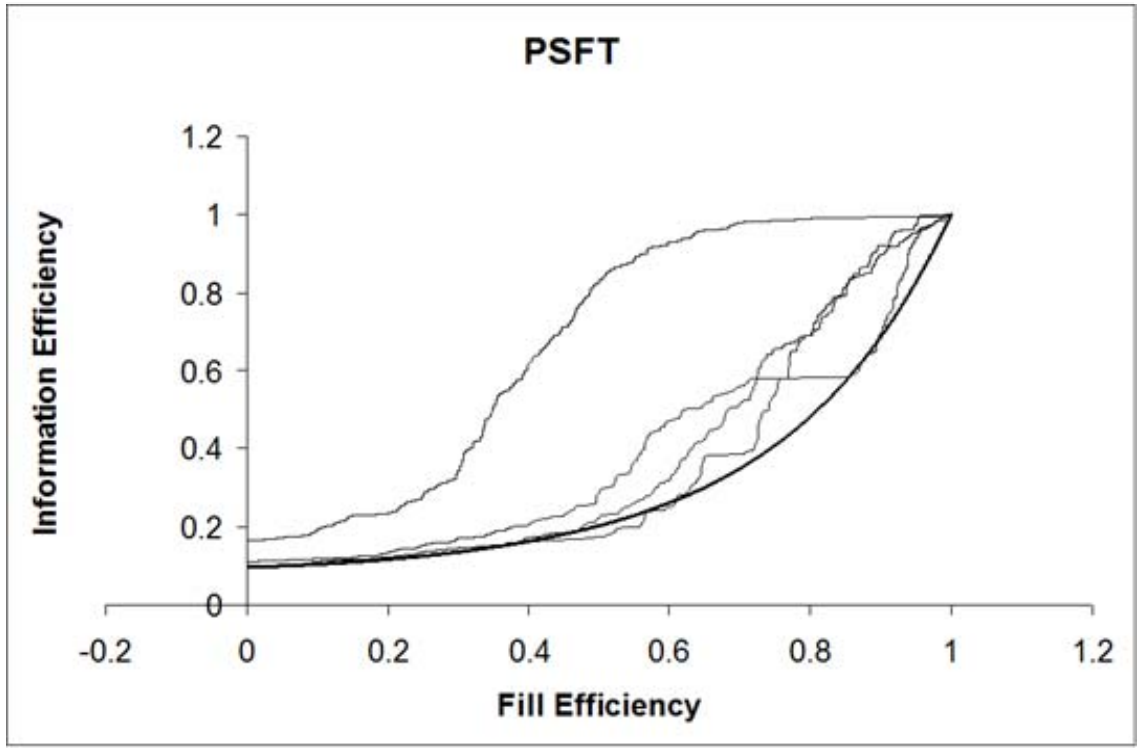

Figure 6: Information Efficiency measurements for PeopleSoft (PSFT).

**Discussion**

We have solved an idealized version of the Maxwell Demon problem and applied it to the definition and measurement of the *information efficiency* of markets. Two fundamental constants common to both problems emerge. In the Maxwell Demon case we calculate the efficiency of an engine driven by information. Because of the similarity between this engine and the whimsical restaurant-driven rocket ship proposed by Douglas Adams[15], we call the constants *Bistro Constants.* We find two regimes, one in which supply/demand for stocks plays an important role and one in which the market orders seem to obey a *Poisson* process.

We propose the following interpretation of the observations:

Early in the trading day, the *information efficiencies* are characterized by a model that accounts for supply/demand in market orders. In fact, at early times in the trading day, the Unfilled to Filled Model fits the data quantitatively well. For very early times, before many orders have been filled, the probabilities for long and short orders are ½ each and the process resembles a *Poisson* Process. As orders are filled, however, the process deviates from a *Poisson* Process.

Late in the trading day traders are pressured to unwind their positions. The observations indicate that a Poisson Process in which long/short market orders enter the market each with probability ½ characterizes the data. The behaviour of the market in this period resembles a *Poisson* Process.

The middle of the day is a transition period for the market. During this period the market moves from a model in which supply/demand dominates to a market characterized by a *Poisson* Process.

The authors would like to thank Seth Myers for his diligent downloading and collection of data.

Correspondence and requests for materials should be addressed to R.D.J. (e-mail: Roger.Jones@Complexica.com).